\documentclass[11pt,a4paper]{revtex4-1}

\usepackage[utf8]{inputenc}
\usepackage[english]{babel}
\usepackage{amsmath}
\usepackage{amsfonts}
\usepackage{amssymb}
\usepackage{graphicx}
\usepackage{mathtools}
\usepackage{bbold}

\def\eq#1{Eq.~\eqref{#1}}
\def\fig#1{Fig.~\ref{#1}}

\newcommand{\barM}[1]{\ensuremath{\mathbf{#1}}}
\renewcommand{\v}[1]{\ensuremath{\mathbf{#1}}}
\newcommand{\pd}[2]{\frac{\partial #1}{\partial #2}} 

\begin{document}

\title{Traction force microscopy on soft elastic substrates:\\
a guide to recent computational advances}
\author{Ulrich S. Schwarz}
\author{J\'er\^ome R.D. Soin\'e}
\affiliation{Institute for Theoretical Physics and Bioquant, Heidelberg University, \\
Philosophenweg 19, 69120 Heidelberg, Germany}

\date{\today}

\begin{abstract}
The measurement of cellular traction forces on soft elastic substrates has become
a standard tool for many labs working on mechanobiology. Here we review the
basic principles and different variants of this approach. In general, the extraction
of the substrate displacement field from image data and the reconstruction procedure
for the forces are closely linked to each other and limited by the presence of
experimental noise. We discuss different strategies to reconstruct cellular
forces as they follow from the foundations of elasticity theory, including 
two- versus three-dimensional, inverse versus direct and linear versus non-linear approaches. 
We also discuss how biophysical models can improve force reconstruction and comment on practical issues
like substrate preparation, image processing and the availability of software for traction force microscopy. 
\end{abstract}

\maketitle

\section{Introduction}

Over the last two decades, it has become apparent that mechanical forces play a central role for cellular decision-making, leading to the emerging field of mechanobiology \cite{mammoto_mechanobiology_2013,iskratsch_appreciating_2014}. In order to understand how forces impact cellular processes, it is essential to measure them with high spatiotemporal resolution and to correlate them either statistically or causally with the cellular process of interest. The most common approach is to measure forces at the cell-matrix interface.
This field has grown rapidly over the last years and has become to be known as \textit{traction force microscopy} (TFM).
Using this approach, it has been shown e.g.\ that cellular traction often correlates with the size of adhesion contacts \cite{balaban_force_2001,tan_cells_2003,goffin_focal_2006,Prager-Khoutorsky_fibroblast_2011,trichet_evidence_2012}
but also that this correlation depends on the growth history of the adhesion contact under consideration \cite{stricker_spatiotemporal_2011,oakes_stressing_2014}. For most tissue cell types, 
high extracellular stiffness correlates with large traction forces
and large cell-matrix adhesion contacts. These large contacts
are thought to not only ensure higher mechanical stability, but also to reflect increased signaling activity.
This leads to a stiffness-sensitive response of cells, e.g.\ during 
cell spreading and migration \cite{pelham_cell_1997,lo_cell_2000} or stem cell differentiation \cite{McBeath_cell_2004,engler_matrix_2006,trappmann_extracellular-matrix_2012,Wen_interplay_2014}. 
While TFM has become a standard tool in many labs working on mechanobiology, in practise the details of its implementation vary significantly and the development of new approaches is moving forward at a very fast pace.

From a general point of view, forces are not an experimentally directly accessible quantity but have to be infered from the fact that they create some kind of motion. Despite the fact that this motion can follow different laws depending on the details of the system under consideration (e.g.\ being elastic or viscous), a force measurement essentially requires to monitor some kind of dynamics. This is illustrated best with a linear elastic spring. Here force is defined as $F=kx$, with spring constant $k$ and displacement $x$. Without a measurement of $x$, no statement on $F$ would be possible ($k$ is a constant that can be obtained from a calibration experiment). In order to measure $x$, the reference state $x=0$ has to be known, and therefore one typically needs a relaxation process to determine the absolute value of $x$. Thus even seemingly static situations require some dynamical measurement. Another instructive example is the stress acting over a fictitious surface inside a static but strained elastic body. In order to measure this stress directly, in principle one has to cut the surface open and to introduce a strain gauge that measures forces by the movement of a calibrated spring. Alternatively one needs to use
a model that allows one to predict this stress from an elastic calculation.

\begin{figure}[h]
\begin{center}
\includegraphics[width=\textwidth]{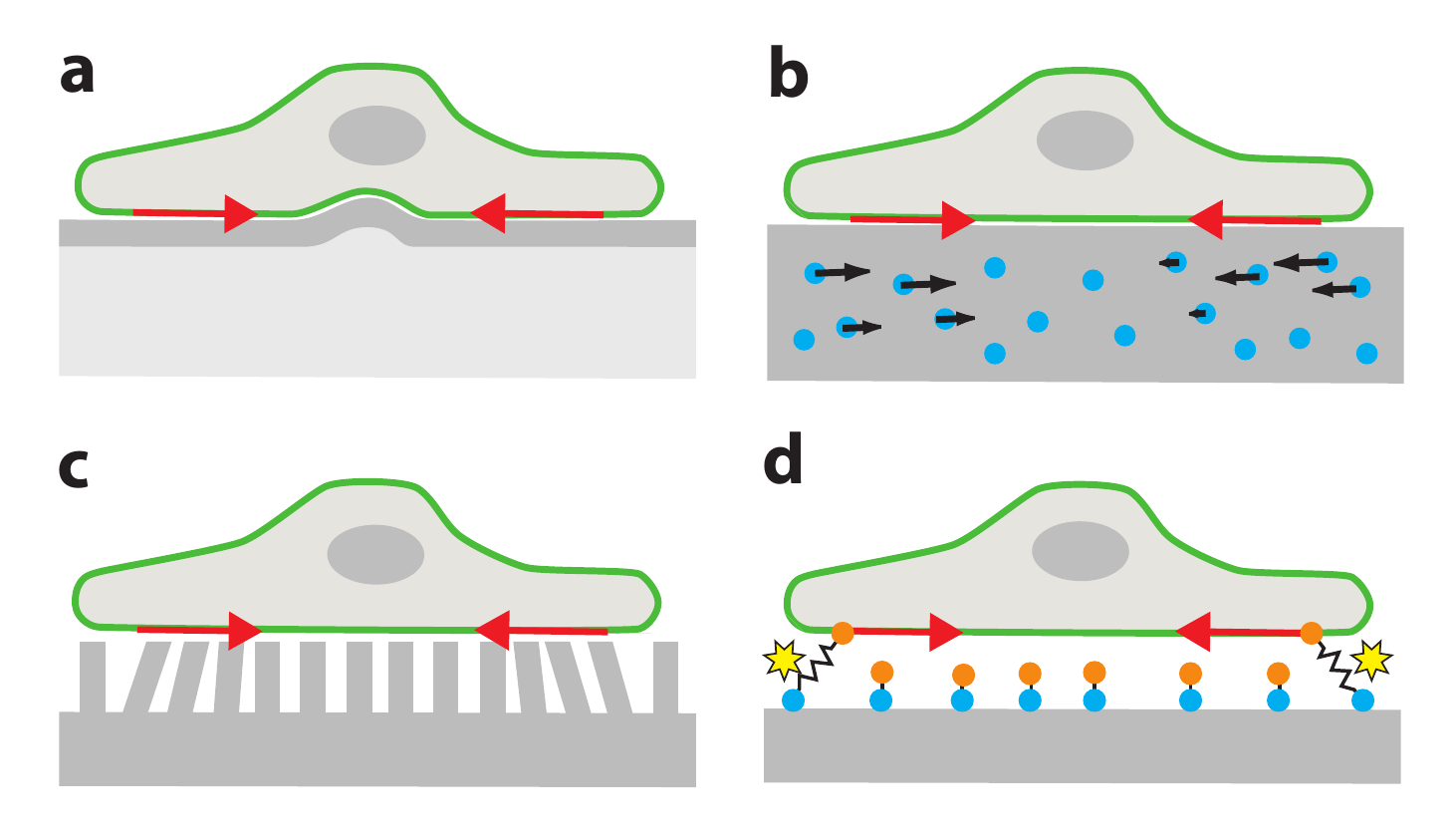}
\end{center}
\caption{Schematic representation of different setups for traction force microscopy.
(a) Thin films buckle under cell traction, therefore this setup is difficult to
evaluate quantitatively. (b) The standard setup are thick films with embedded marker beads as reviewed here.
The substrate deformation field can be extracted with image processing and has to be deconvoluted
to obtain the cellular traction field. (c) Pillar arrays are local strain gauges and do not
require any deconvolution; however, they also present
topographical and biochemical patterns to cells. (d) Fluorescent stress sensors
typically use the relative movement of two molecular domains connected by a calibrated elastic
linker to create a fluorescent signal, e.g.\ by F{\"o}rster resonance energy transfer
(FRET) or by quenching.}
\label{fig_cartoon}
\end{figure}

In summary, each direct measurement of cellular forces has to start with the identification of a suitable strain gauge. Thus a helpful
classification of the wide field of TFM can be introduced by considering the different ways in which a strain gauge can be incorporated in
a cell culture setup (\fig{fig_cartoon}). The most obvious way to do this is to replace the glass or plastic dishes of cell culture by an elastic system that can deform under cell forces. Early attempts to do so used thin elastic sheets, which buckle under cellular traction and thus provide an immediate visual readout (\fig{fig_cartoon}a) \cite{harris_silicone_1980}. However, due to this non-linear response, it is difficult to evaluate these experiments quantitatively. Therefore
this assay was first improved by using thin silicone films under tension \cite{dembo_imaging_1996} and then thick polyacrylamide (PAA) films
that do not buckle but deform smoothly under cell traction (\fig{fig_cartoon}b) \cite{dembo_stresses_1999}. Today
the use of thick films made of different materials is a standard approach in many mechanobiology labs. Fiducial markers can be embedded into these substrates and their movement can be recorded to extract a displacement field. Solving the inverse problem of elasticity theory, cellular traction forces can be calculated from these data \cite{dembo_stresses_1999,butler_traction_2002,schwarz_calculation_2002,sabass_high_2008,plotnikov_high-resolution_2014}. An interesting alternative to solving the inverse problem is the direct method that constructs the stress tensor by a direct mapping from a
strain tensor calculated from the image data \cite{franck_three-dimensional_2007,maskarinec_quantifying_2009,bar-kochba_fast_2015}. Here we will review these methods that are based on the experimental setup shown in \fig{fig_cartoon}b.

A simple alternative to TFM on soft elastic substrates is the use of pillar arrays, where forces are decoupled in an array of local strain gauges (\fig{fig_cartoon}c) \cite{tan_cells_2003,saez_is_2005,roure_force_2005,trichet_evidence_2012,ghassemi_cells_2012}. Pillars can be microfabricated from many different materials, including elastomeres like polydimethylsiloxane (PDMS) or solid material like silicium, as long as they have a sufficiently high aspect ratio to deform under cellular traction. One disadvantage of this approach is that cells are presented with topographical cues and that their adhesion sites grow on laterally restricted islands, making this system fundamentally different from unconstrained adhesion on flat substrates. Moreover it has recently been pointed out that substrate warping might occur if the base is made from the same 
elastic material, thus care has to be taken to correctly calibrate these systems \cite{schoen_probing_2010}.

A very promising alternative to macroscopically large elastic strain gauges is the use of molecular force sensors (\fig{fig_cartoon}d) \cite{grashoff_measuring_2010,stabley_visualizing_2012,morimatsu_molecular_2013,Zhang_DNA-based_2014,Blakely_DNA-based_2014,Liu_nanoparticle_2014,cost_how_2014}. Such a sensor typically consists of 
two molecular domains connected by a calibrated elastic linker. In the example for an extracellular
sensor shown in \fig{fig_cartoon}d, the distal domain is bound to a gold dot on the substrate that
quenches the cell-bound domain and
fluorescence ensues as the linker is stretched by cellular forces \cite{Liu_nanoparticle_2014}.
For intracellular sensors, one can use F{\"o}rster resonance energy transfer (FRET), which
means that fluorescence decreases as the linker is stretched \cite{grashoff_measuring_2010}. 
Fluorescent stress sensors give a direct readout of molecular forces, but for several reasons
one has to be careful when interpreting these signals.
First the effective spring constant of the elastic linker might depend on the
local environment in the cell, even if calibrated in a single-molecule force spectroscopy
experiment. Second the fluorescent signal is a sensitive function of domain separation and
relative orientation, thus a direct conversion into force can be problematic. Third it is difficult to control the number of engaged 
sensors, thus the fluorescent signal cannot easily be integrated over a larger region.
Fourth the molecular stress sensor reads out only part of the force at work in the cellular structure
of interest (e.g.\ the adhesion contact). Therefore fluorescent stress sensors are
expected to complement but not to replace traditional TFM in the future.

One advantage of fluorescent stress sensors over soft elastic substrates and pillar assays
is that they can be more easily adapted to force measurements in tissue, for example in 
developmental systems with fast and complicated cell rearrangements, although the same
issues might apply as discussed above for single cells. Recently, however, it has been shown that 
macroscopic oil droplets can be used to monitor forces during developmental processes \cite{Campas_quantifying_2014}.
In principle, also subcellular structures such as focal adhesions, stress fibers, mitochondria or nuclei can be used
as fiducial markers for cell and tissue deformations \cite{Wang_mechanical_2001,Hu_intracellular_2003}.
One disadvantage of this approach however is that subcellular structures are usually highly dynamic and can exhibit their own
modes of movement, thus not necessarily following the overall deformation of the cell. Nevertheless
conceptually and methodologically these approaches are similar to traditional TFM and also work
in the tissue context.

Another important subfield of TFM is estimating internal forces from cell traction using the concept of force balance. This concept has been
implemented both for forces between few cells \cite{Liu_mechanical_2010,Maruthamuthu_Cell-ECM_2011} and for forces within laterally extended cell monolayers \cite{tambe_collective_2011,tambe_monolayer_2013}.
In the latter case (\textit{monolayer stress microscopy}), one assumes that the cell monolayer behaves like a thin elastic film
coupled to the underlying matrix by stresses (alternatively one can assume coupling by strain \cite{moussus_intracellular_2014,tambe_comment_2014}). Combined with a negative pressure that represents the
effect of actomyosin contractility, the physics of thin elastic films is now
increasingly used to describe forces of cell monolayers in general \cite{edwards_force_2011,mertz_scaling_2012,mertz_cadherin-based_2013,rausch_polarizing_2013}.
Recently single cell and monolayer approaches for internal force reconstruction have been combined by tracking each cell inside a monolayer \cite{Ng_mapping_2014}.
For single cells, the combination of modeling and TFM has recently been advanced to estimate the tensions in the whole set of stress fibers
within cells on pillar arrays \cite{pathak_structural_2012} and soft elastic substrates \cite{Soine_model-based_2015}. For the latter case an actively contracting cable network constructed from image data has been employed to model contractility in the set of stress fibers within U2OS cells. 

Despite the many exciting developments in the large field of TFM, the most commonly used setup to measure cellular forces is traction force microscopy on soft elastic substrates \cite{cesa_micropatterned_2007,wang_cell_2007,style_traction_2014}. Here we review the underlying principles and recent advances with a special focus
on computational aspects. For the following, it is helpful to classify the different approaches in this field. Because cells become rather flat 
in mature adhesion on elastic substrates, traditionally only tangential deformations have been considered (2D TFM).
More recently, tracking of bead movements in all three dimensions has been used to 
reconstruct also z-direction forces (3D TFM) \cite{franck_three-dimensional_2007,maskarinec_quantifying_2009,hur_live_2009,delanoe-ayari_4d_2010,legant_multidimensional_2013}.
For both 2D and 3D TFM, one further has to differ between linear and non-linear procedures.
The central quantity in this context is strain $\epsilon$, which is defined as relative deformation and therefore dimensionless.
If the substrate is sufficiently stiff or the cell sufficiently weak to result in strain values much smaller than unity ($\epsilon \ll 1$), one is in the linear regime and can work with the small strain approximation \cite{dembo_stresses_1999,butler_traction_2002,schwarz_calculation_2002,sabass_high_2008,plotnikov_high-resolution_2014}.
The standard approach to estimating forces is the solution of the inverse problem of linear elasticity theory (inverse TFM).
In the linear case, one can use the Green's function formalism that leads to very fast and efficient algorithms for force reconstruction using inverse procedures,
both for the standard case of thick substrates and for substrates of finite thickness (GF-TFM) \cite{merkel_cell_2007,alamo_spatio-temporal_2007,del_alamo_three-dimensional_2013}. If in contrast one is in the regime
of large strain ($\epsilon \gg 1$), Green's functions cannot be used. One way to
deal with this problem is the use of the finite element method (FEM-TFM)
in a non-linear formulation. Moreover FEM-TFM can also be used
in a linear formulation \cite{yang_determining_2006,legant_measurement_2010}.
For both small and large strain, alternatively the stress can in principle be constructed directly from the displacement (direct TFM) \cite{franck_three-dimensional_2007,maskarinec_quantifying_2009,toyjanova_high_2014,bar-kochba_fast_2015}.

In all approaches used, an important issue is the role of noise on the force reconstruction.
Most TFM-approaches use an inversion of the elastic problem to calculate forces from displacement.
However, this is an ill-posed problem in the sense that due to the long-ranged
nature of elasticity, the calculated traction patterns are highly sensitive to
small variations in the measured displacement field and thus solutions might be ambiguous in the
presence of noise \cite{schwarz_calculation_2002}. Also non-conforming discretization of the problem can cause
ill-posedness, e.g.\ if the mesh is chosen too fine compared to the mean distance of measured bead displacements.
In order to avoid ambiguous solutions, a regularization procedure has to be employed in one way or the other,
e.g.\ by filtering the image data or by adding additional constraints to the force estimation \cite{sabass_high_2008}.
In model-based TFM (MB-TFM), this problem can be avoided if the model is sufficiently limiting  \cite{Soine_model-based_2015}. In direct TFM, care has to be taken how to calculate the derivatives from noisy data \cite{bar-kochba_fast_2015}.
Traction reconstruction with point forces (TRPF) can be considered to be a variant of MB-TFM,
but requires regularization if one uses many point forces \cite{schwarz_calculation_2002,sabass_high_2008}.

\begin{table}
\begin{center}
  \begin{tabular}{| l | l | p{8cm}|}
    \hline
    Abbreviation & Meaning & Comment \\ \hline \hline
    2D TFM & two-dimensional TFM & standard approach using only tangential deformations and forces \\ \hline
    3D TFM & three-dimensional TFM & tracking of 3D deformations allows one to reconstruct 3D forces \\ \hline \hline
    iTFM & inverse TFM & standard approach solving the ill-posed inverse problem of elasticity \\ \hline
    dTFM & direct TFM & if superior 3D image quality is available, the stress tensor can be calculated directly from the strain tensor \\ \hline \hline
    FEM-TFM & Finite element method-based TFM & FEM can be used for iTFM in particular for large strains or complicated geometries \\ \hline
    GF-TFM & Green's function-based TFM & for small strains and simple geometries, Green's functions can be used for iTFM \\ \hline
    BEM & boundary element method & iTFM in real space \\ \hline
    FTTC & Fourier transform traction cytometry & iTFM in Fourier space \\ \hline
    Reg-FTTC & regularized FTTC & FTTC with regularization to deal with noise \\ \hline
    HR-TFM & high resolution TFM & Reg-FTTC with differently coloured marker beads \\ \hline \hline
    TRPF & traction reconstruction with point forces & reconstruction of point-like forces rather than stresses \\ \hline
    MB-TFM & model-based TFM & to avoid regularization and to infer forces beyond traction forces, models can complement TFM \\ \hline
  \end{tabular}
\end{center}
\label{list_variants}
\caption{Abbreviations for different variants of traction force microscopy (TFM) on flat elastic substrates as discussed in this review.
The corresponding references are given in the main text. Regularization is required to deal with experimental noise.
The first five entries in the list exist in both linear and non-linear versions. The other entries are typically used with a linear substrate model.}
\end{table}

In Tab. I we list the different variants of TFM on soft elastic substrates and their respective abbreviations.
In the following we will review recent advances in this field along the lines of this classification.
We start with an introduction into TFM on soft elastic substrates, including an explanation of inverse TFM versus direct TFM.
We next discuss  GF-TFM, FEM-TFM and MB-TFM in more detail. We then
discuss some practical issues arising for TFM-users, namely experimental protocols, 
image processing and software for force reconstruction. We close with discussion and outlook.

\section{Basic principles of elasticity theory and TFM}

The reconstruction of traction forces on soft elastic substrates requires continuum elasticity theory to describe substrate deformations as a consequence of forces being applied to the substrate surface \cite{Landau1983,Holzapfel2000}. Forces applied to the boundary of elastic solids are called traction forces $\mathbf{t}$ and are measured in units of force per area, $Pa = N/m^2$. They lead to substrate deformations which in general are described by
the deformation gradient tensor $\mathbf{F}$, which is the Jacobian of the coordinate transformation from the
undeformed state $\v{x}$ to the deformed state $\v{x'}$:
\begin{equation}
\barM{F}=\begin{pmatrix}
\pd{x'}{x}&\pd{x'}{y}&\pd{x'}{z}\\
\pd{y'}{x}&\pd{y'}{y}&\pd{y'}{z}\\
\pd{z'}{x}&\pd{z'}{y}&\pd{z'}{z}
\end{pmatrix}\ .
\end{equation}
Alternatively one can work with the displacement vector field $\v{u}=\v{x'}-\v{x}$:
\begin{equation}
\barM{F}=\v{1}+(\nabla \otimes \v{u})^T\ .
\end{equation}
Here $\otimes$ is the dyadic product that maps two vectors into a tensor
($(\mathbf{a} \otimes \mathbf{b})_{ij} = a_i b_j$) and
$\nabla$ the gradient operator in the undeformed (Lagrangian) frame.

Both $\mathbf{F}$ and $\v{u}$ are defined at any point in the body and therefore
are spatial fields. They have to be constructed from the image data
and then can be used to derive measures for local changes in distances
and angles during deformation. For this purpose it is useful to define several types of non-linear
strain tensors, in particular the Green-Lagrange tensor $\mathbf{E}$ and
the left Cauchy-Green tensor $\mathbf{B}$:
\begin{equation}
\barM{E}=\frac{1}{2} \left( \barM{F}^T\cdot \barM{F}- \mathbb{1} \right), \quad
\barM{B}=\barM{F} \barM{F}^T\ .
\label{nl_strain}
\end{equation}
In terms of the displacement vector field, the Green-Lagrange tensor can be written as
\begin{equation}
\barM{E} = \frac{1}{2} \left( \left(\nabla \otimes \v{u} \right)+\left(\nabla \otimes \v{u} \right)^T
+\left(\nabla \otimes \v{u} \right) \left(\nabla \otimes \v{u} \right)^T \right)
\end{equation}
which again shows its non-linear nature.
Linearization for small strains gives the linear strain tensor:
\begin{equation}
\barM{\epsilon}=\frac{1}{2}\left(\left(\nabla\otimes\v{u}\right)+\left(\nabla\otimes\v{u}\right)^T\right)\ .
\label{eq_strain_tensor}
\end{equation}
In components this equation reads
\begin{equation}
\epsilon_{ij} = \frac{1}{2} \left( \pd{u_i}{x_j} + \pd{u_j}{x_i} \right)\ .
\label{eq_strain_tensor_comp}
\end{equation}
The linear strain tensor is non-dimensional and describes the relative length changes in different directions.
For simplicity, for the following we assume small strains and thus proceed with the linear strain tensor
(\textit{geometrical linearity}); later we will come back to the large deformation case.

If in addition to small strains one can also assume that the substrate material has a linear and
isotropic constitutive relation (\textit{material linearity}), then a linear relation exists between the strain tensor $\barM{\epsilon}$
and the stress tensor $\barM{\sigma}$ describing the forces acting over internal surfaces:
\begin{equation}
\sigma_{ij} = \frac{E}{1+\nu} \left( \epsilon_{ij} + \frac{\nu}{1-2 \nu} \epsilon_{ll} \delta_{ij} \right)
\label{eq_linear_mapping}
\end{equation} 
where summation over repeated indicies is implied. This equation can be inverted to
\begin{equation}
	\epsilon_{ij} = \frac{1}{E}\left[ \left(1+\nu \right)\sigma_{ij} - \nu \sigma_{ll} \delta_{ij} \right]\ .
\label{eq_linear_mapping_inverted}
\end{equation}
Here $E$ and $\nu$ represent the two elastic constants of the linear and isotropic substrate.
For experiments with tissue cells like fibroblasts, one typically uses a Young's modulus $E \approx 10$ kPa;
for weaker cells like neurons, one typically uses $E \approx 100$ Pa.
The Poisson's ratio $\nu \approx 0.45$ for PAA-films \cite{Kandow_polyacrylamide_2007}
and $\nu \approx 0.5$ for PDMS-films \cite{cesa_micropatterned_2007}. Thus both
material systems are close to being incompressible ($\nu = 0.5$).

The balance of internal and body forces, $\nabla \barM{\sigma} = \v{f}$,
can be written as partial differential equation for the displacement vector field, the Lam\'e equation:
\begin{equation}
\frac{E}{2 (1+\nu)} \Delta \mathbf{u} + \frac{E}{2 (1+\nu)(1-2\nu)} \nabla (\nabla \cdot  \mathbf{u}) = \v{f}\ .
\label{eq_lame}
\end{equation} 
Due to its linearity, the general solution of \eq{eq_lame} can be constructed from the Green's function,
which is the solution for the external force $\v{f}$ being a point force (represented mathematically by the delta function).
For example, a point force solution for infinite elastic space can be calculated (\textit{Kelvin solution}) \cite{Landau1983}.
This solution decays like $1/(E r)$ with Young's modulus and distance $r$ (long-ranged decay)
as the Lam\'e equation \eq{eq_lame} is similar to the Laplace equation.
The angular parts of the Green's function are determined mainly by the Poisson's ratio $\nu$.

In TFM, one typically has $\v{f}=\v{0}$ for the body forces and deformations arise from 
traction forces on the boundaries. However, due to the linearity of \eq{eq_lame},
a propagator can be calculated for point traction forces that in the following we also
denote as Green's function. Most importantly for TFM, such Green's functions
are known for the elastic halfspace (\textit{Boussinesq solution}) \cite{Landau1983} and for the elastic layer of finite thickness \cite{merkel_cell_2007,alamo_spatio-temporal_2007,del_alamo_three-dimensional_2013,style_traction_2014}. 
Once such a Green's function is known, the general solution solution follows as convolution integral: 
\begin{equation}
	\mathbf{u} (\mathbf{x}) = \int \barM{G}(\mathbf{x},\mathbf{x'}) \mathbf{t}(\mathbf{x'}) d\mathbf{x'}\ .
	\label{eq_forward_problem}
\end{equation}
The typical film thickness used in cell experiments is between 50 $\mu$m and 80 $\mu$m.
Then the displacement field decays sufficiently fast into the substrate such that
the boundary conditions at the bottom do not matter and the Boussinesq solution can be used.
For weak cell types, even 20 $\mu$m may suffice. For strong cell types and if a thick substrate
does not fit into the microscope, one can use the Green's function for finite substrate thickness.

The basic equations of elasticity theory reviewed here now suggest two ways to construct
forces from displacements. First one can invert the convolution integral from \eq{eq_forward_problem}
as one wants to infer the traction forces $\v{t}$ from the displacement data $\v{u}$ (inverse TFM).
Alternatively one can directly construct the strain tensor $\barM{\epsilon}$ from the displacement
field $\v{u}$ using \eq{eq_strain_tensor_comp} and from this the stress tensor $\barM{\sigma}$
using \eq{eq_linear_mapping}. The traction forces then follow as $\v{t}=\barM{\sigma}\v{n}$
where $\v{n}$ is the normal to the substrate surface (direct TFM). In the large deformation case,
the linear relation from \eq{eq_linear_mapping} has to be replaced by a non-linear material law.
A standard choice is the Neo-Hookean model leading to \cite{toyjanova_high_2014}
\begin{equation}
\sigma = \frac{E}{2 (1+\nu) det(\barM{F})^{5/3}} \left( \mathbf{B}-\frac{1}{3} tr(\mathbf{B}) \mathbb{1} \right)
+ \frac{E}{3 (1-2 \nu)} ( det(\barM{F})-1 ) \mathbb{1}
\label{neo_Hookean}
\end{equation}
where $\mathbf{B}$ is the left Cauchy-Green tensor defined in \eq{nl_strain}.
For small strains, this simplifies to the linear relation from \eq{eq_linear_mapping}. We note that \eq{neo_Hookean} describes 
compressible materials and that the second term diverges in the incompressible limit $\nu=0.5$. For this case multi-field formulations can be employed where incompressibility is treated separately \cite{Holzapfel2000}.

It is essential to note that all TFM-methods suffer from the presence of 
noise in the experimental data. Noise in TFM-experiments has different origins,
including the limited optical resolution of the microscope, the uncertainties in the image processing procedures
and heterogeneities in the substrate material with its embedded marker beads.
In the case of inverse TFM, the inversion of the convolution integral (a
Fredholm integral equation of the 1st kind) in the presence of noise is ill-posed due to the long-ranged ($1/r$) nature of the elastic
Green's function. This means that small changes in the displacement data can lead to large
changes in the reconstructed forces, thus rendering the inverse procedure unstable. In the case of direct TFM,
calculation of the strain tensor
amounts to taking 3D derivatives of noisy data and therefore can be unstable, too.
Therefore very good 3D image quality has to be achieved
and numerical derivatives have to be calculated with
appropriate noise filters, e.g. the optimal-11 differentiation
kernel \cite{bar-kochba_fast_2015}.
For standard data sets it is therefore more feasible to proceed with inverse TFM. Only recently
has it become possible to implement direct TFM based on much improved image processing
procedures. Therefore in the following we first discuss inverse TFM, in particular its
main variants, the Fourier method FTTC, which is a special version of GF-TFM.

\section{Green's function-based TFM and FTTC}

The first quantitative reconstruction method to measure traction patterns with sub-cellular resolution was proposed for pre-stressed silicone sheets \cite{dembo_imaging_1996}. One disadvantage of this approach is that two-dimensional Green's functions scale
as $\ln r$. Because of the divergence with large $r$, boundary effects become even more important than in the 
case of the long-ranged but decaying Green's function with $1/r$-scaling in three dimensions.
However, single cells are isolated systems that cannot generate momentum and therefore to a first order approximation
their force monopole vanishes and they act as force dipoles \cite{schwarz_physics_2013}. This means that the actual decay in
such experiments should scale as $1/r$ rather
than as $\ln r$ even in two dimensions. Nevertheless the two-dimensional decay
of the Green's function is slow and therefore problematic. 
For this reason this technique was adapted to the case of thick substrates \cite{dembo_stresses_1999}, when the Green's function decays
as $1/r^2$ for the force dipole. With this faster decay, the inversion procedure becomes more stable.

Due to the tensor nature of elasticity theory the Green's function $\barM{G}$ (either for a halfspace or a finite thickness film) is actually a $3 \times 3$ matrix, which accounts for 3D substrate displacements. In most cases it is sufficient and much more simple to record substrate deformations only in lateral directions (2D), however, in this case it is a mandatory requirement that the substrate material is nearly incompressible, which is fulfilled by a Poisson's ratio of $\nu \approx 0.5$. Only under this circumstances the matrix entries in lateral and vertical directions decouple.
We then have $\barM{G}(\mathbf{x},\mathbf{x'})=\barM{G}(\mathbf{x}-\mathbf{x'})$ with
\begin{equation}
\barM{G}(\mathbf{x}) = \frac{(1+\nu)}{\pi E r^3}
\begin{pmatrix}
(1-\nu) r^2+\nu x^2 & \nu x y \\
\nu x y	& (1-\nu) r^2+\nu y^2
\end{pmatrix}.
\label{2d_boussinesq}
\end{equation}
Here $r = \sqrt{x^2 + y^2}$ is the 2D distance. Note that $\barM{G}(\mathbf{x})$ can be inverted everywhere ($\det{G} > 0$) except at the singularity at $\mathbf{x}=\mathbf{0}$. This singularity however is effectively removed by the integration procedure in the convolution integral from \eq{eq_forward_problem}.

The inverse problem can be solved with Green's functions either in real space or in Fourier space.
In real space, this is commonly done using the boundary element method (BEM) \cite{dembo_imaging_1996,dembo_stresses_1999,sabass_high_2008}.
The boundary element method needs in addition to measured substrate deformations the shape of the cell area, which gets spatially discretized by triangulation. The traction field gets thereby represented by a interpolation scheme based on a finite set of nodal traction values (similar to the Galerkin method in FEM). Then \eq{eq_forward_problem} can be written as a linear algebraic equation, which represents a standard boundary element method. To invert the equation a priory information about expected traction field needs to be incorporated.
This is usually implemented by introducing Tikhonov regularization. Here, an additional regularization term is added to the corresponding optimization problem that replaces the inversion of the elastic equation. This regularization term is weightened by a \textit{regularization parameter} $\lambda$ that determines the strength of regularization. In principle there are various choices of regularization terms possible which are characterized by their property to enhance or suppress specific traction distributions. Because cells on flat substrates may exert forces in different directions at neighboring
adhesions, in contrast to many other application areas for regularization methods
smoothness of the traction field (first-order Tikhonov regularization) is not an appropriate choice. In practise one finds 
that inversion without regularization can lead to very large forces as the algorithm tries
to reflect small details of the noisy displacement field with large forces working against each other.
The most reasonable assumption therefore is that traction forces should not become exceedingly large.
One thus arrives at a minimization problem with zero-order Tikhonov regularization:
\begin{equation}
\v{t} = min_{\v{t}} \left( (\barM{G} \v{t} - \v{u})^2 + \lambda^2 \v{t}^2 \right)
\label{minimization}
\end{equation}
where $\v{u}$ is the experimentally measured displacement and $\barM{G} \v{t}$ is the displacement predicted by the estimated traction (the 
convolution integral is not written out here because in practise it is discretized
and thus becomes a matrix operation). The choice of the regularization parameter
$\lambda$ can be based on Bayesian theory \cite{plotnikov_high-resolution_2014} or determined empirically, for example with the L-curve criterion \cite{hansen_rank-deficient_1998}. The BEM is a
conceptually straightforward but computationally expensive method to reconstruct traction patterns. Its main advantages are that
it can reach a high spatial resolution and that it constrains traction to the segmented cell area.

Alternatively one can work in Fourier space (Fourier transform traction cytometry, FTTC) \cite{butler_traction_2002}. In Fourier space the 
convolution integral \eq{eq_forward_problem} factorizes into a product and then can be inverted to give
\begin{equation}
	\tilde{\mathbf{t}}(\mathbf{k}) = \tilde{\barM{G}}(\mathbf{k})^{-1} \tilde{\mathbf{u}}(\mathbf{k})\ .
	\label{eq_inverse_fourier}
\end{equation}
Here the tilde denotes a Fourier transform. The traction reconstruction is achieved in principle by fast Fourier transform of the displacement field, multiplication with the inverse matrix and fast Fourier transform of the result back into real space. The 2D Green's function from \eq{2d_boussinesq} becomes in Fourier space
\begin{equation}
\tilde{\barM{G}}(\mathbf{k}) = \frac{2 (1+\nu)}{E k^3}
\begin{pmatrix}
(1-\nu) k^2+\nu k_y^2 & - \nu k_x k_y \\
- \nu k_x k_y	& (1-\nu) k^2+\nu k_x^2
\end{pmatrix}\ .
\label{2d_boussinesq_fourier}
\end{equation}
FTTC as suggested initially is not regularized and thus could lead to problems in the presence of noise.
In practise, this issue can be attenuated by smoothing the image data and cutting away fast Fourier modes.
A more rigorous approach however is to introduce a regularization scheme into FTTC (Reg-TFM)
\cite{sabass_high_2008,plotnikov_high-resolution_2014}. \eq{eq_inverse_fourier} is then replaced by
\begin{equation}
	\tilde{\mathbf{t}} = \left( \tilde{\barM{G}}^T \tilde{\barM{G}} + \lambda^2 \mathbb{1} \right)^{-1} \tilde{\barM{G}}^T \tilde{\mathbf{u}}
	\label{eq_inverse_fourier_reg}
\end{equation}
where we have left out the $\mathbf{k}$-dependance for notational simplicity.
For regularization parameter $\lambda = 0$, we get back the unregularized solution \eq{eq_inverse_fourier}.

FTTC benefits from a very fast computation time and that no additional information apart from the measured displacement field
is required. Currently it is the most widely used traction reconstruction technique and we show examples of FTTC in action in \fig{fig2}
and \fig{fig3}.
FTTC has been implemented by many labs and also is available as ImageJ plugin \cite{martiel_measurement_2015}.
When working with it, however, one should keep some caveats in mind.
In general one has to note that the resolution of the final traction field strongly depends on the local resolution of substrate deformations according to the Saint-Vernant's principle, which states that the difference between the displacement fields caused
by traction distributions that are different but have the same highest order force multipolar moment 
decays quickly with the distance from the traction source. Therefore in experiments it is essential to record the substrate
deformation close to the force generating processes to resolve the details of the traction
pattern. In practise this means that the deformations right under the cell body are
most relevant. We next note that according to the Nyquist criterion,
the spatial resolution is mainly determined by the bead density as it can be resolved in the image processing procedures.
Therefore nanobeads are prefered over microbeads, which have the additional
advantange of less disturbing the mechanical properties of the substrate. We also note
that fast Fourier transform requires interpolation onto a regular lattice
and therefore a homogeneous distribution of marker beads is essential. 
We further note that for high resolution TFM, one also should keep track of the depth at which the marker bead images
are taken; then an appropriately modified Green's function in Fourier space has to be used \cite{style_traction_2014,plotnikov_high-resolution_2014}. In general, these comments show again that 
the force reconstruction cannot be separated from the quality of the image data and its processing.

\section{FEM-based TFM}

Standard TFM assumes both geometrical and material linearity. If these assumptions are not justified
anymore, then finite element methods (FEM) can be used. This might be the case for soft substrates
or strong cells, and usually is the case if cells
are embedded in a 3D elastic matrix \cite{bloom_mapping_2008,legant_measurement_2010,rehfeldt_hyaluronic_2012,koch_3d_2012}.
Moreover FEM-TFM can also be used in the linear case \cite{yang_determining_2006,legant_measurement_2010}. Here we briefly comment on some applications of this approach.

We first note that the boundary element method (BEM) introduced early by Dembo and coworkers \cite{dembo_stresses_1999}
can be considered to be some variant of FEM-TFM. However, here only the cell and not the substrate
is discretized. It has been shown that TFM can also be completely based on such a discretized
approach, for example in order to take into account finite substrate thickness \cite{yang_determining_2006}.
Here the agreement between measured and simulated displacements has been optimized using
a FEM-formulation. A mathematically very elegant way to reconstruct traction with FEM has
been introduced by Ambrosi and coworkers \cite{Ambrosi_cellular_2006,Ambrosi_traction_2008}.
The basic idea here is to consider the traction reconstruction as PDE constraint optimization problem.
Thus one searches for the stationary state of a Lagrangian type functional that incorporates the elastic PDE as Lagrangian restriction. This leads to a coupled system of Lam\'{e}-type equations associated with the forward and the adjoint problems. By doing this one prevents costly numeric optimization procedures since the optimization problem is reduced to the solution of a coupled system of PDEs, which can be solved in parallel. One disadvantage of this approach is that it assumes thin substrates and plane stress.

For thick substrates, another FEM-approach has been suggested by Hur and coworkers \cite{hur_live_2009,Hur_roles_2012}.
They avoid the inversion of the elastic equations by considering the elastic problem as mixed boundary value problem (BVP). Here, FEM is used as suitable tool to solve the elastic equations with respect to the Cauchy stress $\sigma$, while measured bead displacements enter the equations as boundary condition. The substrate region is considered as cuboid discretized domain and the equation system that has to be solved is represented by inversion of the constitutive relation from \eq{eq_linear_mapping_inverted}.
The cellular traction at the substrate surface follows as $\mathbf{t} = \sigma \mathbf{n}$.
The approach does not use regularization and therefore might be vulnerable to noise.
The authors validated their method with simulated data contaminated with additive Gaussian noise of different strengths and measured a relative deviation of the root mean square $\Delta t =\left \| t_{sim} - t_{recon} \right \| $ of traction up to $25\%$ for a Gaussian noise standard deviation of 200 nm. They further applied the method to data sets of bovine aortic endothelial cells (BAECs) and found prominent non-zero traction stresses in normal direction. These normal forces have now been measured also in
other approaches that reconstruct z-direction forces \cite{maskarinec_quantifying_2009,delanoe-ayari_4d_2010,legant_multidimensional_2013}. 

Another FEM-based approach was proposed by Legant and coworkers to reconstruct traction force patterns of cells embedded in 3D elastic matrix \cite{legant_measurement_2010}. Compared to the BVP approach by Hur and coworkers, this method utilizes FEM not to solve the problem directly, but to calculate a discretized Green's function based on each individual data set including the segmented and triangulated 3D cell shape. For each facet of the cell surface triangulation the displacement responses for unit loads are calculated with FEM to construct a individual discrete Green's function. Based on the calculated Green's function then traditional traction reconstruction is applied using inversion of the convolution formulation from \eq{eq_forward_problem}. Legant et al. inverted the problem by minimizing a least square estimate with a zero-order Tikhonov with respect to the facet loads similar to the TRPF method. The described method represents a smart hybrid method. While it uses the strength of FEM numerics to calculate PDEs on complex geometries, it also traces the problem back to well-established standard TFM. Compared to the Hur approach this represents a higher degree of flexibility, e.g. beads do not have to be distributed directly at the substrate surface. Because this is a Green's function approach, however, this method is limited to linear elasticity theory after all.

\begin{figure}[h]
\begin{center}
\includegraphics[width=\textwidth]{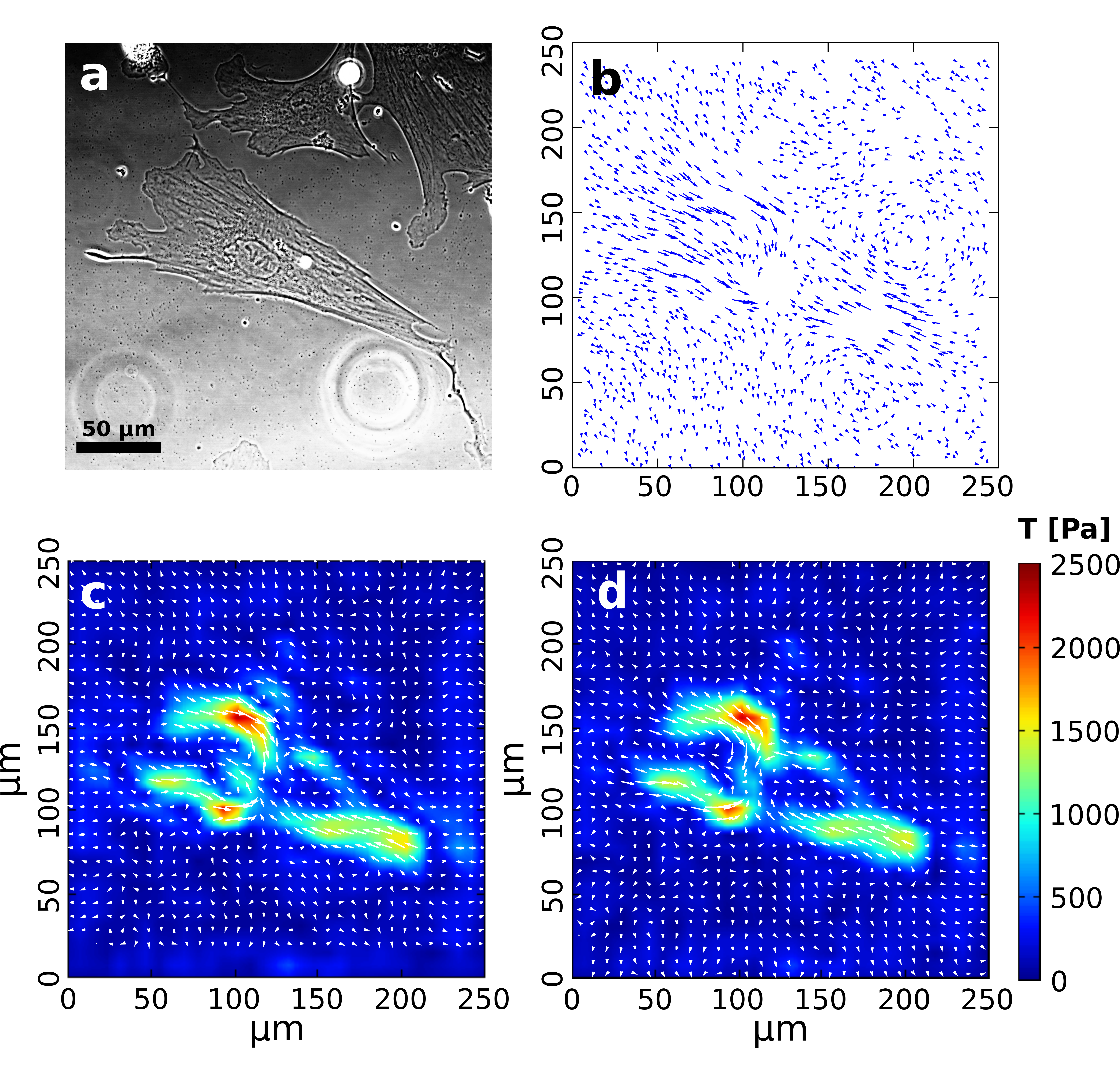}
\end{center}
\caption{Example of force reconstruction with different methods \cite{DissJerome}. (a) Phase contrast image of a cadiac myofibroblast on a $E=15$ kPa PDMS-substrate. (b) Substrate displacement field derived from the displacement of embedded fluorescent marker beads. (c) Reg-FTTC traction reconstruction. (d) FEM-based traction reconstruction. Data courtesy of Merkel group.}
\label{fig2}
\end{figure}

\fig{fig2} depicts a direct comparison of traction reconstruction with standard FTTC and a FEM-based TFM for a cardiac myocyte on a $E=15$ kPa PDMS substrate. The implementation of the FEM-based method used here is described in detail in \cite{DissJerome}. The general idea behind this technique is the replacement of the convolution integral \eq{eq_forward_problem} by the direct elastic PDE, which is solved with respect to a freely choosen traction boundary condition. This boundary condition is parametrized by an interpolation scheme, which consists of a certain number of traction degrees of freedom analogue to the discretization used in FTTC. Traction fields here are reconstructed by repeated calculation of the PDE, while optimizing the traction boundary condition with respect to measured bead displacement fields (beads distributed either on substrate surface or within substrate volume). Note that by replacing the convolution formulation (\eq{eq_forward_problem}) by the FEM-calculation step all constrictions regarding material model and substrate geometry can be overcome, which designates the method to applications like non-planar geometries and non-linear substrate material behavior.
Visual comparison of \fig{fig2}c and d demonstrates that both methods lead to identical traction fields and resolutions. 

\section{Model-based traction force microscopy}

All previous TFM reconstruction methods focused on finding a correct distribution of traction forces with respect to a measured field of substrate deformations. However, this does not tell us how forces are distributed inside the cells. 
This restriction has been lifted for cell monolayers by the introduction of monolayer stress microscopy
which models the cell monolayer as thin elastic sheet \cite{tambe_collective_2011,tambe_monolayer_2013,Ng_mapping_2014}. 
The strategy of combining traction reconstruction with modeling assumptions is very general
and recently has also been applied to the stress fibers in single cells as an example of
model-based TFM \cite{Soine_model-based_2015}.

\begin{figure}[tbp]
\begin{center}
\includegraphics[width=\textwidth]{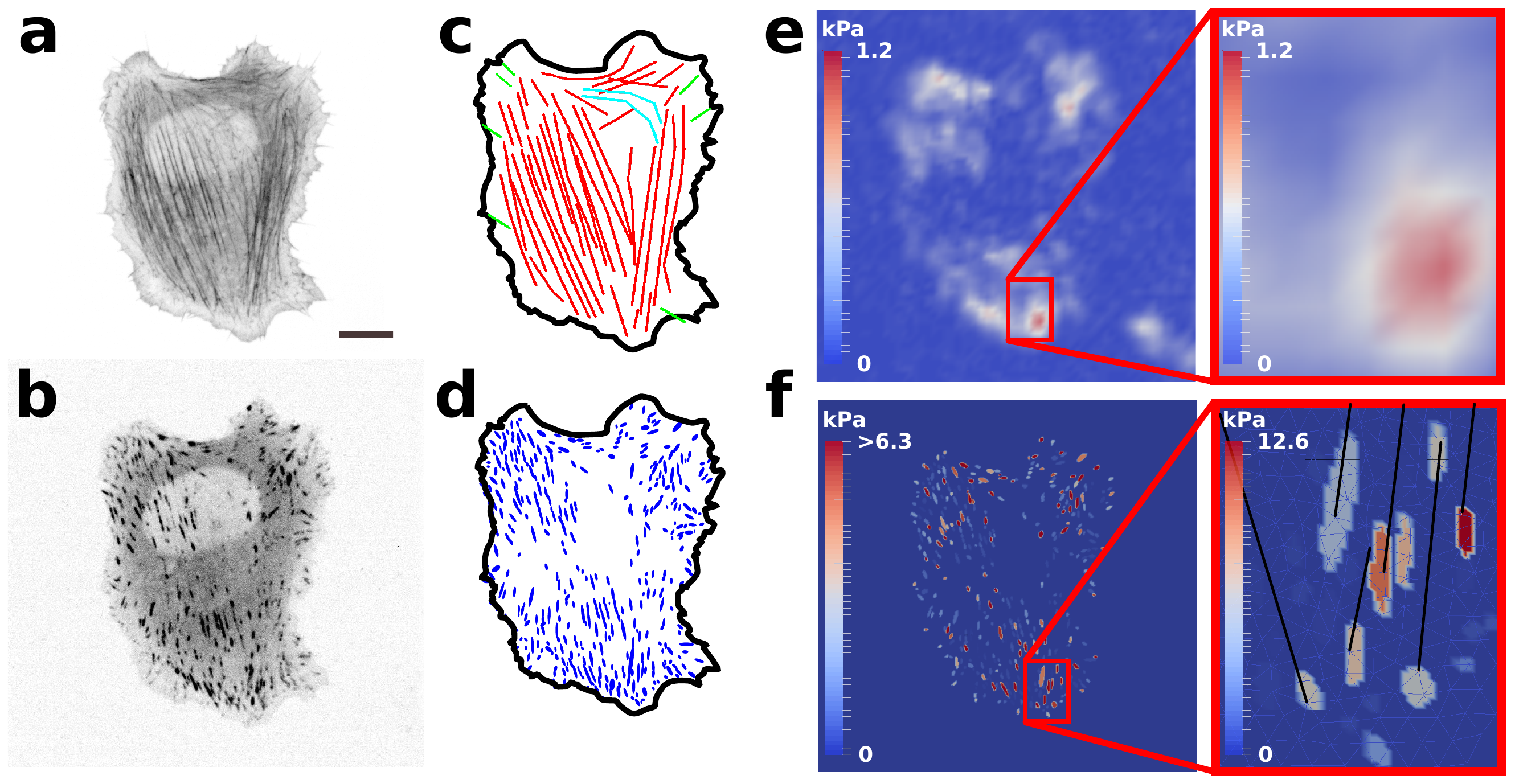}
\end{center}
\caption{Model-based traction force microscopy (MB-TFM) with active cable networks \cite{Soine_model-based_2015}. U2OS-cell on $E=8.4$ kPa PAA substrate. (a) Actin fluorescence image. Scale bar is 10 $\mu$m. (b) Paxillin fluorescence image. (c) Segmented actin stress fibers from actin image. (d) segemented focal adhesions from paxillin image. (e) Reg-FTTC traction force pattern from substrate deformations. (f) MB-TFM traction force pattern based on an active cable cell model. The insets illustrate the different levels of detail obtained from FTTC versus MB-TFM reconstructions. Data courtesy of Gardel group.}
\label{fig3}
\end{figure}

The main idea behind MB-TFM is to combine TFM with quantitative cell modeling. In practice image processing is used to identify prominent mechanical features of the cell based on fluorescence images of focal adhesion proteins and the actin cytoskeleton.
\fig{fig3}a-d shows how these structures can be segmented in live cells (TFM microscopy is only possible in combination with live cell microscopy due to the need to also acquire a reference image). This information is subsequently used to build a individual 2D biophysical model for each cell that represents their mechanical behavior as best as possible while keeping a moderate level of model parameters. For this purpose the cell is represented by a network of contractile active cables. Segmented stress fibers are embedded in the network and after its simulated contraction the resulting forces are transmitted to the substrate at those points where focal adhesion have been segmented before. By doing so, one arrives at a quantitative mechanical cell model with typically 30 to 100 free parameters to optimize, which is much less compared to traditional TFM reconstruction. A comparison of both reconstruction is shown in \fig{fig3}e and f. 

MB-TFM has several advantages compared to standard techniques.
First, the incorporation of the cell model allows us to reveal correlations between substrate deformations and the distribution of intracellular tension. E.g.\ for U2OS cells it could be shown that forces generated in actin stress fibers make up most of the overall traction force, while forces originated from distributed actin networks are in most of the cases negligible. Further it could be validated that different stress fiber types, which have been classified by their assembly mechanism, vary statistically in their contraction strength.
Second, the additional model restriction and according reduction of the parameter space is sufficient to abolish the need for regularization in most cases, while the model filter undertakes the task of a regularization scheme and guarantees that traction field solutions agree with biophysical considerations including e.g. the overall balance of forces. A drawback of this approach is the restriction to a 2D model, the need for very good image data (in order to perform the segmentation and to arrive at a realistic model),
the requirement to image different channels simultaneously (one for the marker beads and
at least two for adhesion and cytoskeletal proteins, respectively) and the high demand on computer power.
With the rapid advances in 3D imaging, in the future this approach might also
be carried into the third dimension. However, although 3D models for contractility of adherent cells
already exist \cite{sen_matrix_2009,ronan_cellular_2013},
it is certainly challenging to acquire the image data needed to parametrize such 3D
models through traction data.

\section{Experimental data and image processing}

We finally comment on some practical issues that arise in TFM.
A typical TFM-experiment proceeds as follows. Polymer monomers and crosslinkers
are mixed at a ratio that corresponds to the desired substrate stiffness.
Then marker beads (commonly used diameters range from $\mu$m down to 40 nm)
are added to this mixture and a drop is deposited on a first coverslip.
A second coverslip is used to sandwich the drop and the
gel is polymerized. The top coverslip is removed again and
the surface coated with extracellular matrix. This culture substrate is then
inserted into the microscope.
Sometimes the beads are centrifuged down towards the substrate surface in order to facilitate image processing.
It it is difficult to dissolve the beads in the substrate material (e.g.\ in PDMS), then
the substrate can be made in a two-step process with the beads deposited on the surface of the first film.
The rigidity of the final substrate can be measured with different techniques, ranging from
microscopic probes like the AFM or magnetic beads to macroscopic probes like film stretching under gravity or rheometers.
Ideally measurements with different setups should give the same results.

Over the years, different protocols have been published for the preparation of
TFM-substrates. An early and very rich resource is volume 83 of Methods in Cell Biology 
edited by Yu-Li Wang and Dennis E. Discher \cite{wang_cell_2007}.
The standard system used for TFM are polyacrylamide (PAA) 
substrates that can easily be tuned over a large stiffness range
\cite{Kandow_polyacrylamide_2007,Moshayedi_mechanosensitivity_2010,Aratyn-Schaus_preparation_2010}.
For high resolution TFM, these protocols can be modified to use differently
colored marker beads \cite{plotnikov_high-resolution_2014}. Surface functionalization
is usually achieved by activating the surface with Sulfo-SANPAH or hydrazine, 
which then can bind ECM-ligands such as collagen or fibronectin. Because acrylamide monomers
are toxic (for example, they react with the amino acid cystein),
care has to be taken that the substrates are indeed completely polymerized
before being used for cell culture. Typically one prepares a stock solution that
can be kept for several years and then prepares working solution on demand.

An alternative system to PAA-substrates
are silicone-oil based substrates (mainly PDMS, that is
commercially available as Sylgard 184) for which also different protocols are available
\cite{cesa_micropatterned_2007,style_traction_2014}. Their high
index of refraction makes them suitable for simultaneous use with
TIRF \cite{Gutierrez_high_2011}. However, they are also more difficult to handle
in regard to inclusion of marker beads, which often are added in a multiple-step process
between two PDMS-layers \cite{style_traction_2014}. Alternatively, the
good patterning properties of PDMS can be used to create some surface
features that can be followed in phase contrast \cite{balaban_force_2001}.
Another issue with PDMS-substrates is that they are very sticky and therefore
hard to handle; at the same time, it is difficult to achieve the kPa-stiffness
needed for cell experiments. Motivated by the widespread use of
TFM-substrates in mechanobiology, recently the difference between PAA- and PDMS-substrates
for stem cell differentiation have been investigated in great detail
\cite{trappmann_extracellular-matrix_2012,Wen_interplay_2014}.

An exciting development in TFM is the combination with micropatterning
approaches, which allow one to normalize cell shape and intracellular
organization \cite{vignaud_directed_2012}. Different techniques
have been developed to pattern soft elastic substrates, including
microcontact printing
\cite{Parker_Directional_2002,Damljanovic_bulk_2005,stricker_Optimization_2010,Rape_regulation_2011}
and deep-ultraviolet illumination of
PAA-substrates \cite{tseng_spatial_2012,oakes_geometry_2014,mandal_cell_2014} as well as lift-off techniques on PDMS-substrates
\cite{hampe_defined_2014}. Combining TFM and cell patterning can
also be accomplished in three dimensions, e.g. using direct laser
writing \cite{klein_elastic_2010,klein_two-component_2011}.

When working with soft elastic substrates, one always has to keep in mind that many different physical 
processes might cause them to deviate from the ideal elastic behaviour assumed in the force reconstruction procedure.
For the soft kPa-substrates required for TFM, care has to be taken to avoid viscous components
(e.g.\ PDMS at 0.1 kPa \cite{trappmann_extracellular-matrix_2012})
or heterogeneities (e.g.\ PAA at unfavorable mixing ratios \cite{Moshayedi_mechanosensitivity_2010})
because both cell behaviour and the force reconstruction procedures are affected by such substrate properties.
If substrates with dissipative elements are used, their viscoelastic properties (creep and relaxation functions)
should be characterized in detail \cite{muller_dissipative_2013}. For the widely used PAA-substrates,
another effect that one has to keep in mind is the possible uptake and release of water
by the deformed hydrogel \cite{Casares_Hydraulic_2015}.

In a TFM experiment, one typically acquires a stack of fluorescence images, including
at least one image for the substrate deformation. At the end of the experiment, one
has to remove the cell (typically by trypsination or by adding similar enzymes, but alternatively by scratching or
by adding calyculin such that the cell rips itself off the substrate) in order to
obtain a reference image. Then the deformation vector field has to be constructed
by image processing and comparison of the deformed and the undeformed images.
There exist two standard solutions plus a number of 
hybrid approaches \cite{style_traction_2014}. In single particle tracking, one uses a low number of fiducial markers
such that one can track them one by one. In correlative tracking,
also known as particle image velocimetry (PIV) or digital image correlation (DIC),
one tracks patterns rather than single particles. Often one uses a combination of
both approaches, e.g. PIV to get a first estimate and then single particle tracking
(possibly augmented by neighborhood features) for a more precise one. The 3D variant of DIC is known under the term digital volume correlation (DVC). In HR-TFM, one can in addition use the correlation between different colors
\cite{sabass_high_2008,plotnikov_high-resolution_2014}.

\begin{table}
\begin{center}
  \begin{tabular}{| p{7cm} |  p{8cm} |}
    \hline
Description & URL \\ \hline \hline
ImageJ plugin FTTC by Qingzong Tseng &
https://sites.google.com/site/qingzongtseng/tfm \\ \hline
Finite thickness TFM from Dufresne group &
http:/doi.org/10.1039/C4SM00264D \\ \hline
Regularization tools by Per Hansen &
http://www.mathworks.com/matlabcentral/ fileexchange/52-regtools \\ \hline
Matlab PIV toolbox by Nobuhito Mori &
http://www.oceanwave.jp/softwares/mpiv/ \\ \hline
Particle tracking code by Daniel Blair and Eric Dufresne &
http://physics.georgetown.edu/matlab/ \\ \hline
Large deformation 3D TFM from Franck group &
http://franck.engin.brown.edu/{\textasciitilde}christianfranck/ FranckLab/Downloads.html \\ \hline
\end{tabular}
\end{center}
\label{tab_software}
\caption{List of helpful software for TFM.}
\end{table}

Over the last years, different software packages have been developed for the
image processing and force reconstruction tasks. In Tab. II we list some of
them for the convenience of the TFM-user. The ImageJ plugin for FTTC
is a ready-to-use solution to implement standard TFM \cite{martiel_measurement_2015}.
The finite thickness software is required when cells are very strong or
the substrate is very thin \cite{style_traction_2014}. The regularization tools by
Hansen can be used for regularized optimization strategies in inverse TFM 
based on optimization, compare \eq{minimization} \cite{hansen_rank-deficient_1998}.
Helpful image processing tools are available also from other fields like 
environmental physics, fluids dynamics, microrheology, single molecule imaging or super resolution microscopy.
A standard choice for correlative tracking is the Matlab PIV toolbox.
Alternatively one can use single particle tracking routines. For
large deformation 3D TFM, a specialized code has been developed
very recently \cite{bar-kochba_fast_2015}.

\section{Conclusion and outlook}

Here we have reviewed recent progress in TFM on soft elastic substrates from
the computational perspective. Our overview shows that this field is moving
at a very fast pace and that many different variants of this approach have
been developed over the last two decades, each with its respective advantages and disadvantages.
Thus the situation is similar to the one for optical microscopy, a field in
which also a lot of progress has been made over the last decades at
many different fronts simultaneously, ranging from super resolution microscopy (STED, PALM, STORM, SIM)
through correlative and fluctuation microscopy to light sheet microscopy (SPIM).
Like in this field, too, for TFM the choice of method depends on the experimental
question one is addressing and in many case a combination of different approaches will
work best.

Our review shows that force reconstruction can not be separated from
data analysis and image processing, in particular due to the noise issue. Irrespective
of the approach used, care has to be taken to deal with the experimental noise
that is always present in the displacement data. Each of the methods 
discussed above includes some kind of regularization, either
implicitly through image filtering or explicitly through some regularization scheme.

For 2D TFM, the standard approach is FTTC and this has become the common
procedure in many labs working in mechanobiology due to its short computing
times. Reg-FTTC makes computation time only slightly longer but introduces
a more rigorous treatment of the noise issue. FTTC can be extended
easily to 3D TFM, but only if the image data is of very good quality. 
The simplest version of 3D TFM implies tracking of marker beads in z-direction
on planar substrates. As such
experiments usually require relatively soft substrates, one typically leaves
the linear domain and large deformation methods have to be used for
force reconstruction. Similar techniques can then be used also in
full 3D TFM, e.g. when cells are encapsulated in hydrogels. Here
however care has to be taken that the cellular environment is indeed
elastic; otherwise a viscoelastic or even plastic theory has to be employed.

The more complex the questions and experiments become that are conducted
in mechanobiology, the more difficult it will get to extract meaningful
correlations and cause-effects relations. We therefore envision
that in the future, such experiments will be increasingly combined
with mathematical models that allow us to extract useful information
from microscopy data in a quantitative manner. Simple examples
discussed above are TRPF and MB-TFM, which use the assumptions
of localization of force transmission to the adhesion contacts (TRPF)
and force generation in the actin cytoskeleton (MB-TFM) to
improve the quality of the data that one can extract from TFM-experiments.
In a similar vein, we expect that in the future, more and more data
will be extracted from microscopy images based on some Bayesian
assumptions that have been validated before by other experimental results.
Another very exciting development is the combination of TFM with
fluorescent stress sensors, which complement it with molecular information
and which can be more easily used in a tissue context.

\begin{acknowledgments}
The authors acknowledge support by the BMBF-programm 
MechanoSys and by the Heidelberg cluster of excellence CellNetworks
through its program for emerging collaborative topics.
We thank Christoph Brand for critical reading and help with \fig{fig3}
for MB-TFM. We thank Nils Hersch, Georg Dreissen, Bernd Hoffmann and Rudolf Merkel for
the data used in \fig{fig2} and Jonathan Stricker, Patrick Oakes and Margaret Gardel
for the data used in \fig{fig3}. We apologize to all authors whose work we could not cite
for space reasons.
\end{acknowledgments}


\begin{thebibliography}{10}

\bibitem{mammoto_mechanobiology_2013}
Tadanori Mammoto, Akiko Mammoto, and Donald~E. Ingber.
\newblock Mechanobiology and {Developmental} {Control}.
\newblock {\em Annual Review of Cell and Developmental Biology} 29 (2013) 27--61.

\bibitem{iskratsch_appreciating_2014}
Thomas Iskratsch, Haguy Wolfenson, and Michael~P. Sheetz.
\newblock Appreciating force and shape - the rise of mechanotransduction in
  cell biology.
\newblock {\em Nature Reviews Molecular Cell Biology} 15 (2014) 825--833.

\bibitem{balaban_force_2001}
Nathalie~Q. Balaban, Ulrich~S. Schwarz, Daniel Riveline, Polina Goichberg, Gila
  Tzur, Ilana Sabanay, Diana Mahalu, Sam Safran, Alexander Bershadsky, Lia
  Addadi, and Benjamin Geiger.
\newblock Force and focal adhesion assembly: a close relationship studied using
  elastic micropatterned substrates.
\newblock {\em Nat Cell Biol} 3 (2001) 466--472.

\bibitem{tan_cells_2003}
John~L. Tan, Joe Tien, Dana~M. Pirone, Darren~S. Gray, Kiran Bhadriraju, and Christopher~S. Chen.
\newblock Cells lying on a bed of microneedles: {An} approach to isolate
  mechanical force.
\newblock {\em Proceedings of the National Academy of Sciences} 100 (2003) 1484.

\bibitem{goffin_focal_2006}
Jerome~M. Goffin, Philippe Pittet, Gabor Csucs, Jost~W. Lussi, Jean-Jacques
  Meister, and Boris Hinz.
\newblock Focal adhesion size controls tension-dependent recruitment of
  alpha-smooth muscle actin to stress fibers.
\newblock {\em The Journal of Cell Biology} 172 (2006) 259--268.

\bibitem{Prager-Khoutorsky_fibroblast_2011}
Masha Prager-Khoutorsky, Alexandra Lichtenstein, Ramaswamy Krishnan, Kavitha
  Rajendran, Avi Mayo, Zvi Kam, Benjamin Geiger, and Alexander~D. Bershadsky.
\newblock Fibroblast polarization is a matrix-rigidity-dependent process
  controlled by focal adhesion mechanosensing.
\newblock {\em Nat Cell Biol} 13 (2011) 1457--1465.

\bibitem{trichet_evidence_2012}
Léa Trichet, Jimmy Le~Digabel, Rhoda~J Hawkins, Sri Ram~Krishna Vedula, Mukund
  Gupta, Claire Ribrault, Pascal Hersen, Raphaël Voituriez, and Benoît
  Ladoux.
\newblock Evidence of a {Large}-{Scale} {Mechanosensing} {Mechanism} for
  {Cellular} {Adaptation} to {Substrate} {Stiffness}.
\newblock {\em Proceedings of the National Academy of Sciences} 109 (2012) 6933-6938.

\bibitem{stricker_spatiotemporal_2011}
Jonathan Stricker, Yvonne Aratyn-Schaus, Patrick~W. Oakes, and Margaret~L.
  Gardel.
\newblock Spatiotemporal {Constraints} on the {Force}-{Dependent} {Growth} of
  {Focal} {Adhesions}.
\newblock {\em Biophysical Journal} 100 (2011) 2883--2893.

\bibitem{oakes_stressing_2014}
Patrick~W Oakes and Margaret~L Gardel.
\newblock Stressing the limits of focal adhesion mechanosensitivity.
\newblock {\em Current Opinion in Cell Biology} 30 (2014) 68--73.

\bibitem{pelham_cell_1997}
Robert~J. Pelham and Yu-li Wang.
\newblock Cell locomotion and focal adhesions are regulated by substrate
  flexibility.
\newblock {\em Proceedings of the National Academy of Sciences} 94 (1997) 13661--13665.

\bibitem{lo_cell_2000}
Chun-Min Lo, Hong-Bei Wang, Micah Dembo, and Yu-Li Wang.
\newblock Cell {Movement} {Is} {Guided} by the {Rigidity} of the {Substrate}.
\newblock {\em Biophysical Journal} 79 (2000) 144--152.

\bibitem{McBeath_cell_2004}
Rowena McBeath, Dana~M Pirone, Celeste~M Nelson, Kiran Bhadriraju, and
  Christopher~S Chen.
\newblock Cell {Shape}, {Cytoskeletal} {Tension}, and {RhoA} {Regulate} {Stem}
  {Cell} {Lineage} {Commitment}.
\newblock {\em Developmental Cell} 6 (2004) 483--495.

\bibitem{engler_matrix_2006}
A~Engler, S~Sen, H~Sweeney, and D~Discher.
\newblock Matrix {Elasticity} {Directs} {Stem} {Cell} {Lineage}
  {Specification}.
\newblock {\em Cell} 126 (2006) 677--689.

\bibitem{trappmann_extracellular-matrix_2012}
Britta Trappmann, Julien~E. Gautrot, John~T. Connelly, Daniel G.~T. Strange,
  Yuan Li, Michelle~L. Oyen, Martien A.~Cohen Stuart, Heike Boehm, Bojun Li,
  Viola Vogel, Joachim~P. Spatz, Fiona~M. Watt, and Wilhelm T.~S. Huck.
\newblock Extracellular-matrix tethering regulates stem-cell fate.
\newblock {\em Nature Materials} 11 (2012) 642--649.

\bibitem{Wen_interplay_2014}
Jessica~H. Wen, Ludovic~G. Vincent, Alexander Fuhrmann, Yu~Suk Choi, Kolin~C.
  Hribar, Hermes Taylor-Weiner, Shaochen Chen, and Adam~J. Engler.
\newblock Interplay of matrix stiffness and protein tethering in stem cell
  differentiation.
\newblock {\em Nature Materials} 13 (2014) 979--987.

\bibitem{harris_silicone_1980}
AK~Harris, P~Wild, and D~Stopak.
\newblock Silicone rubber substrata: a new wrinkle in the study of cell
  locomotion.
\newblock {\em Science} 208 (1980) 177 --179.

\bibitem{dembo_imaging_1996}
M.~Dembo, T.~Oliver, A.~Ishihara, and K.~Jacobson.
\newblock Imaging the traction stresses exerted by locomoting cells with the
  elastic substratum method.
\newblock {\em Biophysical Journal} 70 (1996) 2008--2022.

\bibitem{dembo_stresses_1999}
Micah Dembo and Yu-Li Wang.
\newblock Stresses at the {Cell}-to-{Substrate} {Interface} during {Locomotion}
  of {Fibroblasts}.
\newblock {\em Biophysical Journal} 76 (1999) 2307--2316.

\bibitem{butler_traction_2002}
James~P. Butler, Iva~Marija Tolić-Nørrelykke, Ben Fabry, and Jeffrey~J.
  Fredberg.
\newblock Traction fields, moments, and strain energy that cells exert on their
  surroundings.
\newblock {\em American Journal of Physiology - Cell Physiology} 282 (2002) C595--C605.

\bibitem{schwarz_calculation_2002}
U.S. Schwarz, N.Q. Balaban, D.~Riveline, A.~Bershadsky, B.~Geiger, and S.A.
  Safran.
\newblock Calculation of {Forces} at {Focal} {Adhesions} from {Elastic}
  {Substrate} {Data}: {The} {Effect} of {Localized} {Force} and the {Need} for
  {Regularization}.
\newblock {\em Biophysical Journal} 83 (2002) 1380--1394.

\bibitem{sabass_high_2008}
B~Sabass, ML~Gardel, CM~Waterman, and US~Schwarz.
\newblock High resolution traction force microscopy based on experimental and
  computational advances.
\newblock {\em Biophysical Journal} 94 (2008) 207--220.

\bibitem{plotnikov_high-resolution_2014}
Sergey~V. Plotnikov, Benedikt Sabass, Ulrich~S. Schwarz, and Clare~M. Waterman.
\newblock High-{Resolution} {Traction} {Force} {Microscopy}.
\newblock In Jennifer C. Waters {and}~Torsten Wittman, editor, {\em Methods in
  {Cell} {Biology}}, volume 123 (2014) 367--394.

\bibitem{franck_three-dimensional_2007}
C.~Franck, S.~Hong, S.~A. Maskarinec, D.~A. Tirrell, and G.~Ravichandran.
\newblock Three-dimensional {Full}-field {Measurements} of {Large}
  {Deformations} in {Soft} {Materials} {Using} {Confocal} {Microscopy} and
  {Digital} {Volume} {Correlation}.
\newblock {\em Experimental Mechanics} 47 (2007) 427--438.

\bibitem{maskarinec_quantifying_2009}
Stacey~A. Maskarinec, Christian Franck, David~A. Tirrell, and Guruswami
  Ravichandran.
\newblock Quantifying cellular traction forces in three dimensions.
\newblock {\em Proceedings of the National Academy of Sciences of the United
  States of America} 106 (2009) 22108--22113.

\bibitem{bar-kochba_fast_2015}
E.~Bar-Kochba, J.~Toyjanova, E.~Andrews, K.-S. Kim, and C.~Franck.
\newblock A {Fast} {Iterative} {Digital} {Volume} {Correlation} {Algorithm} for
  {Large} {Deformations}.
\newblock {\em Experimental Mechanics} 55 (2015) 261--274.

\bibitem{saez_is_2005}
Alexandre Saez, Axel Buguin, Pascal Silberzan, and Benoît Ladoux.
\newblock Is the {Mechanical} {Activity} of {Epithelial} {Cells} {Controlled}
  by {Deformations} or {Forces}?
\newblock {\em Biophysical Journal} 89 (2005) L52--L54.

\bibitem{roure_force_2005}
Olivia~du Roure, Alexandre Saez, Axel Buguin, Robert~H. Austin, Philippe
  Chavrier, Pascal Siberzan, and Benoit Ladoux.
\newblock Force mapping in epithelial cell migration.
\newblock {\em Proceedings of the National Academy of Sciences of the United
  States of America} 102 (2005) 2390--2395.

\bibitem{ghassemi_cells_2012}
Saba Ghassemi, Giovanni Meacci, Shuaimin Liu, Alexander~A. Gondarenko, Anurag
  Mathur, Pere Roca-Cusachs, Michael~P. Sheetz, and James Hone.
\newblock Cells test substrate rigidity by local contractions on submicrometer
  pillars.
\newblock {\em Proceedings of the National Academy of Sciences} 109 (2012) 5328--5333.

\bibitem{schoen_probing_2010}
Ingmar Schoen, Wei Hu, Enrico Klotzsch, and Viola Vogel.
\newblock Probing {Cellular} {Traction} {Forces} by {Micropillar} {Arrays}:
  {Contribution} of {Substrate} {Warping} to {Pillar} {Deflection}.
\newblock {\em Nano Letters} 10 (2010) 1823--1830.

\bibitem{grashoff_measuring_2010}
Carsten Grashoff, Brenton~D. Hoffman, Michael~D. Brenner, Ruobo Zhou, Maddy
  Parsons, Michael~T. Yang, Mark~A. McLean, Stephen~G. Sligar, Christopher~S.
  Chen, Taekjip Ha, and Martin~A. Schwartz.
\newblock Measuring mechanical tension across vinculin reveals regulation of
  focal adhesion dynamics.
\newblock {\em Nature} 466 (2010) 263--266.

\bibitem{stabley_visualizing_2012}
Daniel~R. Stabley, Carol Jurchenko, Stephen~S. Marshall, and Khalid~S. Salaita.
\newblock Visualizing mechanical tension across membrane receptors with a
  fluorescent sensor.
\newblock {\em Nature Methods} 9 (2012) 64--67.

\bibitem{morimatsu_molecular_2013}
Masatoshi Morimatsu, Armen~H. Mekhdjian, Arjun~S. Adhikari, and Alexander~R.
  Dunn.
\newblock Molecular {Tension} {Sensors} {Report} {Forces} {Generated} by
  {Single} {Integrin} {Molecules} in {Living} {Cells}.
\newblock {\em Nano Letters} 13 (2013) 3985--3989.

\bibitem{Zhang_DNA-based_2014}
Yun Zhang, Chenghao Ge, Cheng Zhu, and Khalid Salaita.
\newblock {DNA}-based digital tension probes reveal integrin forces during
  early cell adhesion.
\newblock {\em Nature Communications} 5 (2014).

\bibitem{Blakely_DNA-based_2014}
Brandon~L. Blakely, Christoph~E. Dumelin, Britta Trappmann, Lynn~M. McGregor,
  Colin~K. Choi, Peter~C. Anthony, Van~K. Duesterberg, Brendon~M. Baker,
  Steven~M. Block, David~R. Liu, and Christopher~S. Chen.
\newblock A {DNA}-based molecular probe for optically reporting cellular
  traction forces.
\newblock {\em Nature Methods} 11 (2014) 1229--1232.

\bibitem{Liu_nanoparticle_2014}
Yang Liu, Rebecca Medda, Zheng Liu, Kornelia Galior, Kevin Yehl, Joachim~P.
  Spatz, Elisabetta~Ada Cavalcanti-Adam, and Khalid Salaita.
\newblock Nanoparticle {Tension} {Probes} {Patterned} at the {Nanoscale}:
  {Impact} of {Integrin} {Clustering} on {Force} {Transmission}.
\newblock {\em Nano Letters} 14 (2014) 5539--5546.

\bibitem{cost_how_2014}
Anna-Lena Cost, Pia Ringer, Anna Chrostek-Grashoff, and Carsten Grashoff.
\newblock How to {Measure} {Molecular} {Forces} in {Cells}: {A} {Guide} to
  {Evaluating} {Genetically}-{Encoded} {FRET}-{Based} {Tension} {Sensors}.
\newblock {\em Cellular and Molecular Bioengineering} 8 (2014) 96-105.

\bibitem{Campas_quantifying_2014}
Otger Campas, Tadanori Mammoto, Sean Hasso, Ralph~A. Sperling, Daniel
  O'Connell, Ashley~G. Bischof, Richard Maas, David~A. Weitz, L.~Mahadevan, and
  Donald~E. Ingber.
\newblock Quantifying cell-generated mechanical forces within living embryonic
  tissues.
\newblock {\em Nature Methods} 11 (2014) 183--189.

\bibitem{Wang_mechanical_2001}
Ning Wang, Keiji Naruse, Dimitrije Stamenović, Jeffrey~J. Fredberg,
  Srboljub~M. Mijailovich, Iva~Marija Tolić-Nørrelykke, Thomas Polte, Robert
  Mannix, and Donald~E. Ingber.
\newblock Mechanical behavior in living cells consistent with the tensegrity
  model.
\newblock {\em Proceedings of the National Academy of Sciences} 98 (2001) 7765--7770.

\bibitem{Hu_intracellular_2003}
Shaohua Hu, Jianxin Chen, Ben Fabry, Yasushi Numaguchi, Andrew Gouldstone,
  Donald~E. Ingber, Jeffrey~J. Fredberg, James~P. Butler, and Ning Wang.
\newblock Intracellular stress tomography reveals stress focusing and
  structural anisotropy in cytoskeleton of living cells.
\newblock {\em American Journal of Physiology - Cell Physiology} 285 (2003) C1082--C1090.

\bibitem{Liu_mechanical_2010}
Zhijun Liu, John~L. Tan, Daniel~M. Cohen, Michael~T. Yang, Nathan~J. Sniadecki,
  Sami~Alom Ruiz, Celeste~M. Nelson, and Christopher~S. Chen.
\newblock Mechanical tugging force regulates the size of cell–cell junctions.
\newblock {\em Proceedings of the National Academy of Sciences} 107 (2010) 9944--9949.

\bibitem{Maruthamuthu_Cell-ECM_2011}
Venkat Maruthamuthu, Benedikt Sabass, Ulrich~S. Schwarz, and Margaret~L.
  Gardel.
\newblock Cell-{ECM} traction force modulates endogenous tension at cell–cell
  contacts.
\newblock {\em Proceedings of the National Academy of Sciences} 108 (2011) 4708--4713.

\bibitem{tambe_collective_2011}
Dhananjay~T. Tambe, C.~Corey~Hardin, Thomas~E. Angelini, Kavitha Rajendran,
  Chan~Young Park, Xavier Serra-Picamal, Enhua~H. Zhou, Muhammad~H. Zaman,
  James~P. Butler, David~A. Weitz, Jeffrey~J. Fredberg, and Xavier Trepat.
\newblock Collective cell guidance by cooperative intercellular forces.
\newblock {\em Nature Materials} 10 (2011) 469--475.

\bibitem{tambe_monolayer_2013}
Dhananjay~T. Tambe, Ugo Croutelle, Xavier Trepat, Chan~Young Park, Jae~Hun Kim,
  Emil Millet, James~P. Butler, and Jeffrey~J. Fredberg.
\newblock Monolayer {Stress} {Microscopy}: {Limitations}, {Artifacts}, and
  {Accuracy} of {Recovered} {Intercellular} {Stresses}.
\newblock {\em PLoS ONE} 8 (2013) e55172.

\bibitem{moussus_intracellular_2014}
Michel Moussus, Christelle~der Loughian, David Fuard, Marie Courçon, Danielle
  Gulino-Debrac, Hélène Delanoë-Ayari, and Alice Nicolas.
\newblock Intracellular stresses in patterned cell assemblies.
\newblock {\em Soft Matter} 10 (2014) 2414--2423.

\bibitem{tambe_comment_2014}
Dhananjay~T. Tambe, James~P. Butler, and Jeffrey~J. Fredberg.
\newblock Comment on “{Intracellular} stresses in patterned cell
  assemblies” by {M}. {Moussus} et al., {Soft} {Matter}, 2014, 10, 2414.
\newblock {\em Soft Matter} 10 (2014) 7681--7682.

\bibitem{edwards_force_2011}
Carina~M. Edwards and Ulrich~S. Schwarz.
\newblock Force {Localization} in {Contracting} {Cell} {Layers}.
\newblock {\em Physical Review Letters} 107 (2011) 128101.

\bibitem{mertz_scaling_2012}
Aaron~F. Mertz, Shiladitya Banerjee, Yonglu Che, Guy~K. German, Ye~Xu, Callen
  Hyland, M.~Cristina Marchetti, Valerie Horsley, and Eric~R. Dufresne.
\newblock Scaling of {Traction} {Forces} with the {Size} of {Cohesive} {Cell}
  {Colonies}.
\newblock {\em Physical Review Letters} 108 (2012) 198101.

\bibitem{mertz_cadherin-based_2013}
Aaron~F. Mertz, Yonglu Che, Shiladitya Banerjee, Jill~M. Goldstein, Kathryn~A.
  Rosowski, Stephen~F. Revilla, Carien~M. Niessen, M.~Cristina Marchetti,
  Eric~R. Dufresne, and Valerie Horsley.
\newblock Cadherin-based intercellular adhesions organize epithelial
  cell–matrix traction forces.
\newblock {\em Proceedings of the National Academy of Sciences} 110 (2013) 842--847.

\bibitem{rausch_polarizing_2013}
Sebastian Rausch, Tamal Das, Jerome Soine, Tobias Hofmann, Christian Boehm,
  Ulrich Schwarz, Heike Boehm, and Joachim Spatz.
\newblock Polarizing cytoskeletal tension to induce leader cell formation
  during collective cell migration.
\newblock {\em Biointerphases} 8 (2013) 32.

\bibitem{Ng_mapping_2014}
Mei~Rosa Ng, Achim Besser, Joan~S. Brugge, and Gaudenz Danuser.
\newblock Mapping the dynamics of force transduction at cell-cell junctions of
  epithelial clusters.
\newblock {\em eLife} 3 (2014) e03282.

\bibitem{pathak_structural_2012}
Amit Pathak, Christopher~S. Chen, Anthony~G. Evans, and Robert~M. McMeeking.
\newblock Structural {Mechanics} {Based} {Model} for the {Force}-{Bearing}
  {Elements} {Within} the {Cytoskeleton} of a {Cell} {Adhered} on a {Bed} of
  {Posts}.
\newblock {\em Journal of Applied Mechanics} 79 (2012) 061020--061020.

\bibitem{Soine_model-based_2015}
Jerome R.~D. Soine, Christoph~A. Brand, Jonathan Stricker, Patrick~W. Oakes,
  Margaret~L. Gardel, and Ulrich~S. Schwarz.
\newblock Model-based {Traction} {Force} {Microscopy} {Reveals} {Differential}
  {Tension} in {Cellular} {Actin} {Bundles}.
\newblock {\em PLoS Comput Biol} 11 (2015) e1004076.

\bibitem{cesa_micropatterned_2007}
Claudia~M. Cesa, Norbert Kirchgessner, Dirk Mayer, Ulrich~S. Schwarz, Bernd
  Hoffmann, and Rudolf Merkel.
\newblock Micropatterned silicone elastomer substrates for high resolution
  analysis of cellular force patterns.
\newblock {\em Review of Scientific Instruments} 78 (2007) 034301.

\bibitem{wang_cell_2007}
Yu-Li Wang and Dennis~E. Discher. {\em Methods in
  {Cell} {Biology}}, volume 83 (2007).

\bibitem{style_traction_2014}
Robert~W. Style, Rostislav Boltyanskiy, Guy~K. German, Callen Hyland,
  Christopher~W. MacMinn, Aaron~F. Mertz, Larry~A. Wilen, Ye~Xu, and Eric~R.
  Dufresne.
\newblock Traction force microscopy in physics and biology.
\newblock {\em Soft Matter} 10 (2014) 4047--4055.

\bibitem{hur_live_2009}
Sung~Sik Hur, Yihua Zhao, Yi-Shuan Li, Elliot Botvinick, and Shu Chien.
\newblock Live {Cells} {Exert} 3-{Dimensional} {Traction} {Forces} on {Their}
  {Substrata}.
\newblock {\em Cellular and Molecular Bioengineering} 2 (2009) 425--436.

\bibitem{delanoe-ayari_4d_2010}
H.~Delanoe-Ayari, J.~P. Rieu, and M.~Sano.
\newblock 4d {Traction} {Force} {Microscopy} {Reveals} {Asymmetric} {Cortical}
  {Forces} in {Migrating} {Dictyostelium} {Cells}.
\newblock {\em Physical Review Letters} 105 (2010) 248103.

\bibitem{legant_multidimensional_2013}
Wesley~R. Legant, Colin~K. Choi, Jordan~S. Miller, Lin Shao, Liang Gao, Eric
  Betzig, and Christopher~S. Chen.
\newblock Multidimensional traction force microscopy reveals out-of-plane
  rotational moments about focal adhesions.
\newblock {\em Proceedings of the National Academy of Sciences} 110 (2013) 881--886.

\bibitem{merkel_cell_2007}
Rudolf Merkel, Norbert Kirchgessner, Claudia~M. Cesa, and Bernd Hoffmann.
\newblock Cell {Force} {Microscopy} on {Elastic} {Layers} of {Finite}
  {Thickness}.
\newblock {\em Biophysical Journal} 93 (2007) 3314--3323.

\bibitem{alamo_spatio-temporal_2007}
Juan C.~del Alamo, Ruedi Meili, Baldomero Alonso-Latorre, Javier
  Rodriguez-Rodríguez, Alberto Aliseda, Richard~A. Firtel, and Juan~C.
  Lasheras.
\newblock Spatio-temporal analysis of eukaryotic cell motility by improved
  force cytometry.
\newblock {\em Proceedings of the National Academy of Sciences} 104 (2007) 13343--13348.

\bibitem{del_alamo_three-dimensional_2013}
Juan~C. del Alamo, Ruedi Meili, Begoña Alvarez-Gonzalez, Baldomero
  Alonso-Latorre, Effie Bastounis, Richard Firtel, and Juan~C. Lasheras.
\newblock Three-{Dimensional} {Quantification} of {Cellular} {Traction}
  {Forces} and {Mechanosensing} of {Thin} {Substrata} by {Fourier} {Traction}
  {Force} {Microscopy}.
\newblock {\em PLoS ONE} 8 (2013) e69850.

\bibitem{yang_determining_2006}
Zhaochun Yang, Jeen-Shang Lin, Jianxin Chen, and James H-C. Wang.
\newblock Determining substrate displacement and cell traction fields—a new
  approach.
\newblock {\em Journal of Theoretical Biology} 242 (2006) 607--616.

\bibitem{legant_measurement_2010}
Wesley~R Legant, Jordan~S Miller, Brandon~L Blakely, Daniel~M Cohen, Guy~M
  Genin, and Christopher~S Chen.
\newblock Measurement of mechanical tractions exerted by cells in
  three-dimensional matrices.
\newblock {\em Nat Meth} 7 (2010) 969--971.

\bibitem{toyjanova_high_2014}
Jennet Toyjanova, Eyal Bar-Kochba, Cristina López-Fagundo, Jonathan Reichner,
  Diane Hoffman-Kim, and Christian Franck.
\newblock High {Resolution}, {Large} {Deformation} 3d {Traction} {Force}
  {Microscopy}.
\newblock {\em PLoS ONE} 9 (2014) e90976.

\bibitem{Landau1983}
L.~D. Landau and E.M. Lifschitz.
\newblock {\em Theory of elasticity}, volume~7.
\newblock Akademie-Verlag, 1983.

\bibitem{Holzapfel2000}
Gerhard~A. Holzapfel.
\newblock {\em Nonlinear Solid Mechanics}.
\newblock Wiley, 2006.

\bibitem{Kandow_polyacrylamide_2007}
Casey~E. Kandow, Penelope~C. Georges, Paul~A. Janmey, and Karen~A. Beningo.
\newblock Polyacrylamide {Hydrogels} for {Cell} {Mechanics}: {Steps} {Toward}
  {Optimization} and {Alternative} {Uses}.
\newblock In Yu-Li Wang {and} Dennis~E. Discher, editor, {\em Methods in
  {Cell} {Biology}}, volume~83, pages 29--46.
  Academic Press, 2007.

\bibitem{schwarz_physics_2013}
Ulrich~S. Schwarz and Samuel~A. Safran.
\newblock Physics of adherent cells.
\newblock {\em Reviews of Modern Physics} 85 (2013) 1327--1381.

\bibitem{hansen_rank-deficient_1998}
Per~Christian Hansen.
\newblock {\em Rank-{Deficient} and {Discrete} {Ill}-{Posed} {Problems}:
  {Numerical} {Aspects} of {Linear} {Inversion}}.
\newblock SIAM, 1998.

\bibitem{DissJerome}
J{\'e}r{\^o}me~R.D. Soin{\'e}.
\newblock {\em Reconstruction and Simulation of Cellular Traction Forces}.
\newblock PhD thesis, Heidelberg University, 2014.

\bibitem{martiel_measurement_2015}
Jean-Louis Martiel, Aldo Leal, Laetitia Kurzawa, Martial Balland, Irene Wang,
  Timothée Vignaud, Qingzong Tseng, and Manuel Théry.
\newblock Measurement of cell traction forces with {ImageJ}.
\newblock In Ewa~K. Paluch, editor, {\em Methods in {Cell} {Biology}}, volume
  125, pages 269--287. Academic Press, 2015.

\bibitem{bloom_mapping_2008}
Ryan~J. Bloom, Jerry~P. George, Alfredo Celedon, Sean~X. Sun, and Denis Wirtz.
\newblock Mapping {Local} {Matrix} {Remodeling} {Induced} by a {Migrating}
  {Tumor} {Cell} {Using} {Three}-{Dimensional} {Multiple}-{Particle}
  {Tracking}.
\newblock {\em Biophysical Journal} 95 (2008) 4077--4088.

\bibitem{rehfeldt_hyaluronic_2012}
Florian Rehfeldt, André E.~X. Brown, Matthew Raab, Shenshen Cai, Allison~L.
  Zajac, Assaf Zemel, and Dennis~E. Discher.
\newblock Hyaluronic acid matrices show matrix stiffness in 2d and 3d dictates
  cytoskeletal order and myosin-{II} phosphorylation within stem cells.
\newblock {\em Integrative Biology} 4 (2012) 422--430.

\bibitem{koch_3d_2012}
Thorsten~M. Koch, Stefan Münster, Navid Bonakdar, James~P. Butler, and Ben
  Fabry.
\newblock 3d {Traction} {Forces} in {Cancer} {Cell} {Invasion}.
\newblock {\em PLoS ONE} 7 (2012) e33476.

\bibitem{Ambrosi_cellular_2006}
D.~Ambrosi.
\newblock Cellular {Traction} as an {Inverse} {Problem}.
\newblock {\em SIAM Journal on Applied Mathematics} 66 (2006) 2049--2060.

\bibitem{Ambrosi_traction_2008}
D.~Ambrosi, A.~Duperray, V.~Peschetola, and C.~Verdier.
\newblock Traction patterns of tumor cells.
\newblock {\em Journal of Mathematical Biology} 58 (2008) 163--181.

\bibitem{Hur_roles_2012}
Sung~Sik Hur, Juan C.~del Álamo, Joon~Seok Park, Yi-Shuan Li, Hong~A. Nguyen,
  Dayu Teng, Kuei-Chun Wang, Leona Flores, Baldomero Alonso-Latorre, Juan~C.
  Lasheras, and Shu Chien.
\newblock Roles of cell confluency and fluid shear in 3-dimensional
  intracellular forces in endothelial cells.
\newblock {\em Proceedings of the National Academy of Sciences} 109 (2012) 11110--11115.

\bibitem{sen_matrix_2009}
Shamik Sen, Adam~J. Engler, and Dennis~E. Discher.
\newblock Matrix {Strains} {Induced} by {Cells}: {Computing} {How} {Far}
  {Cells} {Can} {Feel}.
\newblock {\em Cellular and Molecular Bioengineering} 2 (2009) 39--48.

\bibitem{ronan_cellular_2013}
William Ronan, Vikram~S. Deshpande, Robert~M. McMeeking, and J.~Patrick
  McGarry.
\newblock Cellular contractility and substrate elasticity: a numerical
  investigation of the actin cytoskeleton and cell adhesion.
\newblock {\em Biomechanics and Modeling in Mechanobiology} 13 (2013) 417--435.

\bibitem{Moshayedi_mechanosensitivity_2010}
Pouria Moshayedi, Luciano da~F. Costa, Andreas Christ, Stephanie~P. Lacour,
  James Fawcett, Jochen Guck, and Kristian Franze.
\newblock Mechanosensitivity of astrocytes on optimized polyacrylamide gels
  analyzed by quantitative morphometry.
\newblock {\em Journal of Physics: Condensed Matter} 22 (2010) 194114.

\bibitem{Aratyn-Schaus_preparation_2010}
Yvonne Aratyn-Schaus, Patrick~W Oakes, Jonathan Stricker, Stephen~P Winter, and
  Margaret~L Gardel.
\newblock Preparation of complaint matrices for quantifying cellular
  contraction.
\newblock {\em Journal of Visualized Experiments: JoVE} 46 (2010).

\bibitem{Gutierrez_high_2011}
Edgar Gutierrez, Eugene Tkachenko, Achim Besser, Prithu Sundd, Klaus Ley,
  Gaudenz Danuser, Mark~H. Ginsberg, and Alex Groisman.
\newblock High {Refractive} {Index} {Silicone} {Gels} for {Simultaneous}
  {Total} {Internal} {Reflection} {Fluorescence} and {Traction} {Force}
  {Microscopy} of {Adherent} {Cells}.
\newblock {\em PLoS ONE} 6 (2011).

\bibitem{vignaud_directed_2012}
Timothée Vignaud, Laurent Blanchoin, and Manuel Théry.
\newblock Directed cytoskeleton self-organization.
\newblock {\em Trends in Cell Biology} 22 (2012) 671--682.

\bibitem{Parker_Directional_2002}
Kevin~Kit Parker, Amy~Lepre Brock, Cliff Brangwynne, Robert~J. Mannix, Ning
  Wang, Emanuele Ostuni, Nicholas~A. Geisse, Josehphine~C. Adams, George~M.
  Whitesides, and Donald~E. Ingber.
\newblock Directional control of lamellipodia extension by constraining cell
  shape and orienting cell tractional forces.
\newblock {\em The FASEB Journal} 16 (2002) 1195 --1204.

\bibitem{Damljanovic_bulk_2005}
V~Damljanovic, BC~Lagerholm, and K~Jacobson.
\newblock Bulk and micropatterned conjugation of extracellular matrix proteins
  to characterized polyacrylamide substrates for cell mechanotransduction
  assays.
\newblock {\em BioTechniques} 39 (2005) 847--851.

\bibitem{stricker_Optimization_2010}
Jonathan Stricker, Benedikt Sabass, Ulrich~S Schwarz, and Margaret~L Gardel.
\newblock Optimization of traction force microscopy for micron-sized focal
  adhesions.
\newblock {\em Journal of Physics: Condensed Matter} 22 (2010) 194104.

\bibitem{Rape_regulation_2011}
Andrew~D. Rape, Wei-hui Guo, and Yu-li Wang.
\newblock The regulation of traction force in relation to cell shape and focal
  adhesions.
\newblock {\em Biomaterials} 32 (2011) 2043--2051.

\bibitem{tseng_spatial_2012}
Qingzong Tseng, Eve Duchemin-Pelletier, Alexandre Deshiere, Martial Balland,
  Hervé Guillou, Odile Filhol, and Manuel Théry.
\newblock Spatial organization of the extracellular matrix regulates
  cell–cell junction positioning.
\newblock {\em Proceedings of the National Academy of Sciences} 109 (2012) 1506--1511.

\bibitem{oakes_geometry_2014}
Patrick W. Oakes, Shiladitya Banerjee, M.~Cristina Marchetti, and Margaret L.
  Gardel.
\newblock Geometry {Regulates} {Traction} {Stresses} in {Adherent} {Cells}.
\newblock {\em Biophysical Journal} 107 (2014) 825--833.

\bibitem{mandal_cell_2014}
Kalpana Mandal, Irène Wang, Elisa Vitiello, Laura Andreina~Chacòn Orellana,
  and Martial Balland.
\newblock Cell dipole behaviour revealed by {ECM} sub-cellular geometry.
\newblock {\em Nature Communications} 5 (2014).

\bibitem{hampe_defined_2014}
Nico Hampe, Thorsten Jonas, Benjamin Wolters, Nils Hersch, Bernd Hoffmann, and
  Rudolf Merkel.
\newblock Defined 2-{D} microtissues on soft elastomeric silicone rubber using
  lift-off epoxy-membranes for biomechanical analyses.
\newblock {\em Soft Matter} 10 (2014) 2431--2443.

\bibitem{klein_elastic_2010}
Franziska Klein, Thomas Striebel, Joachim Fischer, Zhongxiang Jiang, Clemens~M
  Franz, Georg von Freymann, Martin Wegener, and Martin Bastmeyer.
\newblock Elastic {Fully} {Three}-dimensional {Microstructure} {Scaffolds}
  for {Cell} {Force} {Measurements}.
\newblock {\em Advanced Materials} 22 (2010) 868--871.

\bibitem{klein_two-component_2011}
Franziska Klein, Benjamin Richter, Thomas Striebel, Clemens~M Franz, Georg~von
  Freymann, Martin Wegener, and Martin Bastmeyer.
\newblock Two-{Component} {Polymer} {Scaffolds} for {Controlled}
  {Three}-{Dimensional} {Cell} {Culture}.
\newblock {\em Advanced Materials} 23 (2011).

\bibitem{muller_dissipative_2013}
Christina Müller, Andreas Müller, and Tilo Pompe.
\newblock Dissipative interactions in cell–matrix adhesion.
\newblock {\em Soft Matter} 9 (2013) 6207--6216.

\bibitem{Casares_Hydraulic_2015}
Laura Casares, Romaric Vincent, Dobryna Zalvidea, Noelia Campillo, Daniel
  Navajas, Marino Arroyo, and Xavier Trepat.
\newblock Hydraulic fracture during epithelial stretching.
\newblock {\em Nature Materials} 14 (2015) 343--351.

\end{thebibliography}

\end{document}